\documentclass{an}
\usepackage{graphicx}
\usepackage{times}
\usepackage{fancyhdr}
 \def\hr{\hbox{$^{\rm h}$}}                 
 \def\deg{\hbox{$^\circ$}}                  
 \def\fsec{\hbox{$.\!\!^{\rm s}$}}          
 \def\arcm{\hbox{$^\prime$}}                
 \def\arcs{\hbox{$^{\prime\prime}$}}        
 \def\day{\hbox{$^{\rm d}$}}                
 \def\mm{\hbox{$^{\rm m}$}}                 

\begin{document}

\title{Optical observations of Be/X-ray transient system  KS 1947+300}

\author{ \"U. K{\i}z{\i}lo\u{g}lu, A. Baykal, N. K{\i}z{\i}lo\u{g}lu}
\institute{Physics Dept., Middle East Technical University, Ankara 06531, Turkey. }

\date{Received; accepted; published online}

\abstract{ROTSE-IIId 
observations of the Be/X-ray transient system KS 1947+300 obtained between
September 2004 and December 2005 make it possible to study
the correlation between optical and X-ray activity. The optical outburst of
0.1 mag was accompanied by an increase in X-ray flux in 2004 observations. 
Strong correlation between the optical and X-ray light curves
suggests that neutron star directly accretes from the outflowing  material
of Be star. The nearly zero time lag between X-ray and optical light curves
suggests a heating of the disk of Be star by X-rays. 
No optical brightening and X-ray enhancement was seen in 2005
observations. There is no indication of the orbital modulation in the
optical light curve.
}

\correspondence{nil@astroa.physics.metu.edu.tr}

\maketitle

\section{Introduction}
The transient X-ray source KS 1947+300 was discovered in 1989 through observations of the GC 2023+338
field (Borozdin et al. 1990). After its discovery Grankin et al.
(1991) and Goranskij et al. (1991) identified an optical counterpart which shows 
H$\alpha$ emission line in its spectrum. Negueruela et al. (2003) derived a
 spectral type B0Ve for the optical companion and a distance of $\sim$ 10 kpc.
 Considering
 this distance, the peak X-ray luminosity during the 2000-01 outburst was 
$\sim 2\,10^{37}$ erg/s (typical for Be/X-ray transients during the 
Type II outburst).
The equivalent width of
H$\alpha$ line remained almost the same during their one year
 observation period.
They mentioned that the occurrence of several weaker X-ray outbursts has not been
 reflected in any change in the intensity of H$\alpha$ line.
They explained this fact with a low inclination angle of the system.
Galloway et al. (2004), analyzing their observations obtained with the
 instruments aboard RXTE (the ASM, PCA, HEXTE), found that neutron star revolves in a nearly circular
orbit (e = 0.033) with orbital period 40.415 d and projected
 semimajor axis 137\,lt-s, based on the modulations of neutron star's pulse period
of $18.7\,\mathrm{s}$ (Chakrabarty et al. 1995) during the 2000-01 outburst. 
They reported an increase in neutron star period indicating that the angular
momentum was effectively transferred from the accreted material
on to the neutron star which required an extended disk.
They did not find eclipses in their data. 
Tsygankov$\&$Lutovinov (2005) estimated the magnetic field strength
of the pulsar as $2.5\,10^{13}$ G and distance to the binary
as $9.5\pm1.1$ kpc. This system occupies the region of 
Be/X systems in the P$_{spin}$- P$_{orb}$ diagram (Reig et al.
2004). 

\section{Observations and Results}

Our CCD observations were obtained between MJD 53255 (September 7, 2004)
 and MJD
53718 (December 14, 2005) using 45 cm ROTSE-IIId robotic telescope located at
Bak{\i}rl{\i}tepe, Turkey.
It operates without filters and has a wide passband which peaks at
550 nm (Akerlof et al., 2003).
ROTSE magnitudes were calculated by comparing all the field stars to
USNO 2.0 R-band catalog.
About 1000 CCD frames which have an exposure time of 5\,s
were analyzed following the procedure
 described in K{\i}z{\i}lo\u{g}lu et al. (2005) and Baykal et al. (2005).
This system has right ascension and declination of
 $19\hr 49\mm 30\fsec5$ and $+30\deg 12\arcm 24\arcs$, respectively.
\begin{figure*}
\resizebox{\hsize}{!} 
 {\includegraphics[angle=-90,scale=2]{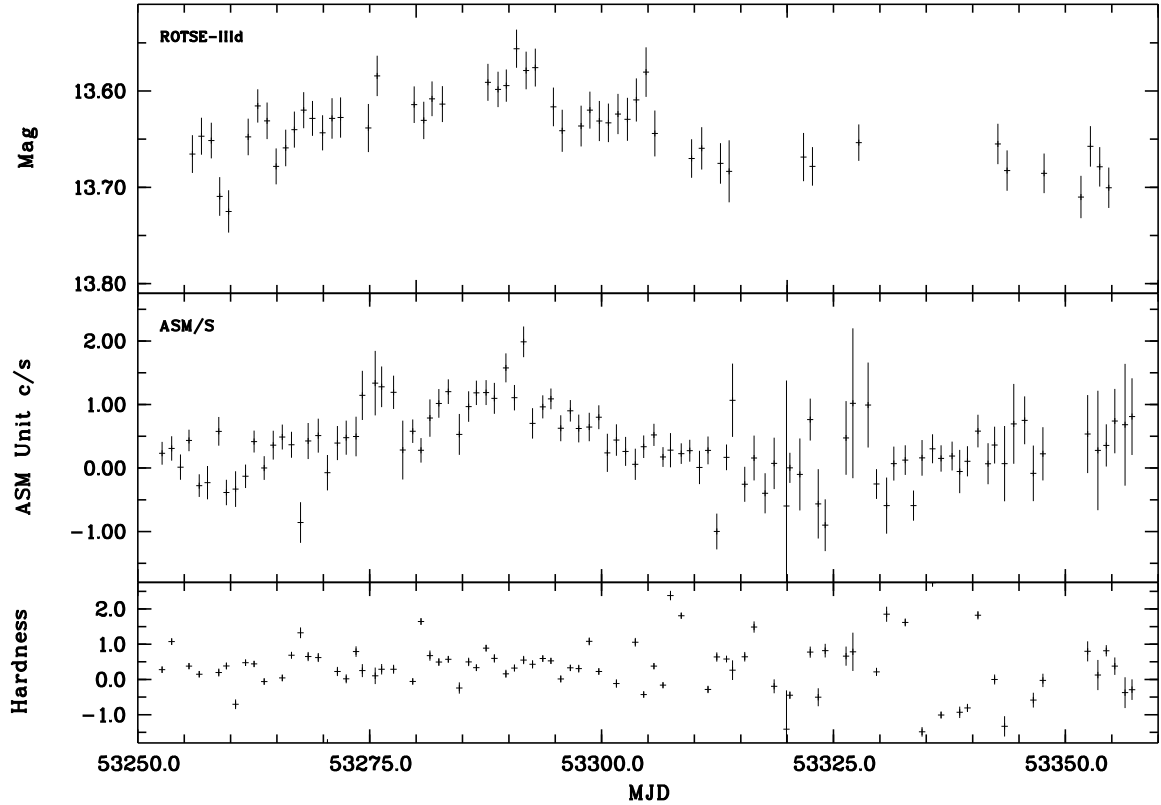}}
\caption{Upper panel shows ROTSE-IIId daily average light curve 
        of KS1949+300 for Sep - Dec 2004 period
        and middle panel represents X-ray light curve taken with RXTE/ASM
        (daily average of 1.5-12.0 keV energy band) for the same period.
        Lower panel is the hardness ratio of RXTE/ASM 3.0-5.0 KeV band
        to 5.0-12.0 KeV band.}
\label{figlabel}
\end{figure*}
\begin{figure*}
\resizebox{\hsize}{!}
{\includegraphics[angle=-90]{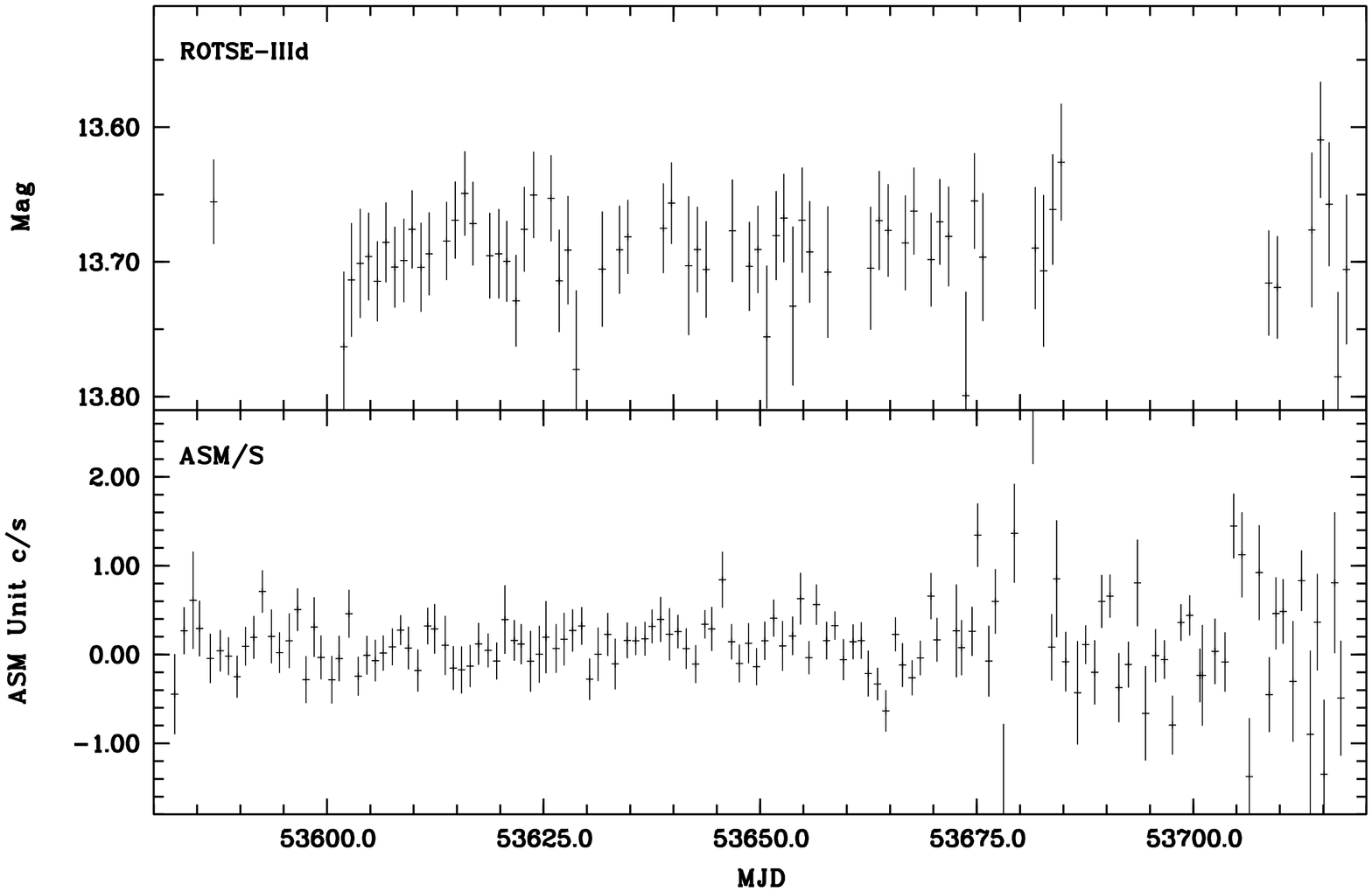}}
\caption{ The same as Fig. 1 but for Jul - Dec 2005 interval.}
\label{figlabel}
\end{figure*}

Figures 1 and 2 show the long time optical observations of
Be/X-ray transient system KS 1947+300 for the first time.
Fig. 1  shows the optical outburst like behavior of this system
for the data obtained in 2004. Fig. 2 shows the optical data for the year
 2005. On the same figures the X-ray light curves in ASM Unit c/s
(daily average of
1.5- 12 keV energy band) of this system which was detected with the All-Sky-Monitor
(ASM) aboard RXTE are plotted. An increase of 0.1 mag in optical magnitude
is seen in Fig. 1 accompanied by an increase in X-ray flux. Both curves show
a similar trend. A change in the measured magnitude of Be star is
 reflected in the change in X-ray flux.
This similarity between optical and X-ray light curves is not seen clearly when
only the low energy band of ASM data (1.5-3.0 keV) was considered.
During optical and X-ray excess the hardness ratio of the higher energy bands
(3.0-5.0 and 5.0-12.0 keV) did not show any notable change.
In 2005, both light curves are seen nearly constant during the
optical observation period.
There is no significant magnitude change in the optical
light curve and likewise no enhancement in the X-ray light curve.

A classical  correlation analysis (Oppenheim and Schaffer 1975) was used to 
test for correlated/time-lagged changes in X-ray and optical light curves.
The classical correlation function is defined as

\[
 CF(\tau) = {<[a(t) - \bar a][b(t + \tau) - \bar b]> \over \sigma_{a}\sigma_{b} }
\]
 
where $<...>$ is the expectation value of the two time series lagged with
respect to  each other. $\sigma_{a}$ and $\sigma_{b}$ are the standard 
deviations of each time series. The time series and the CF are  properly 
normalized such that at $\tau = 0$ the autocorrelation of a single 
time series is unity.
Fig.3 is plot of the cross-correlation between optical and X-ray
light curves.
A very strong correlation exists between the optical
 and X-ray light curves for the 2004 data. 
The correlation which is centered on a nearly zero time-lag,
starts $\sim 20$ days before and lasts $\sim 35$ days. Correlated X/optical
data suggests that neutron star directly accretes from the outflowing 
material of Be star.
The 2005 data indicates that both optical and X-ray time series are not
correlated for all time-lags.
\begin{figure}
\resizebox{\hsize}{!} 
{\includegraphics[angle=-90]{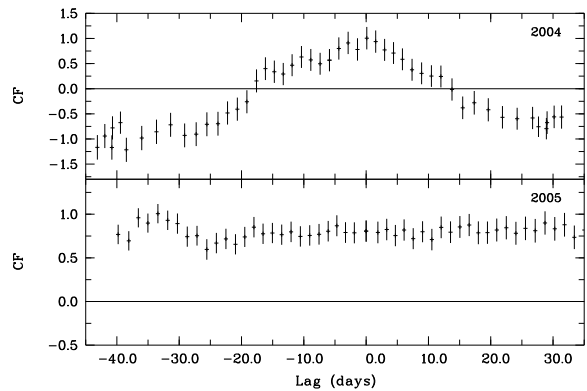}}
\caption{Cross correlation of the ROTSE-IIId and ASM light 
         curves for Sep - Dec 2004 (upper panel) and  
         Jul - Dec 2005 observations (lower panel)}
\label{figlabel}  
\end{figure} 

In Fig.4 we present timing analysis for the ROTSE-IIId light curve. 
Barycentric corrections were made to the time series by using JPL DE200 
ephemerides prior to the timing analysis.  The power spectrum build from all
ROTSE-IIId data (Sep 2004 - Dec 2005) by using Scargle method (Scargle 1982) 
did not show any pronounced frequency. The power at the frequency 1 d$^{-1}$ is
the effect of nightly observations as expected. We could not find
any indication of the  orbital period modulation of KS1947+300.
However there seems to be a red noise behavior for the frequencies lower than
0.07 d$^{-1}$. The slope of the red noise spectrum is $\sim \nu^{-1.6}$,
where $\nu$ is the frequency in d$^{-1}$.

Middle panel in the Fig. 4 is the power spectrum calculated for the 2004
ROTSE-IIId data. It is the same time interval that a correlation between X-ray
and optical light curves is seen. Again no apparent modulation can be
recognized. Although orbital modulation is not expected in such a short 
observation span, other short time scale modulations are also absent. 
No significant period is detected.
An indication of a red noise feature seen in 2004 - 2005 data is also present
for the frequencies lower than $\sim 0.07$ d$^{-1}$.
The power spectrum for the 2005 observations shown in the lower panel of 
Fig. 4 also does not show any specific periodic behavior. The red noise
feature is absent.

\begin{figure}
\resizebox{\hsize}{!}
{\includegraphics[angle=-90]{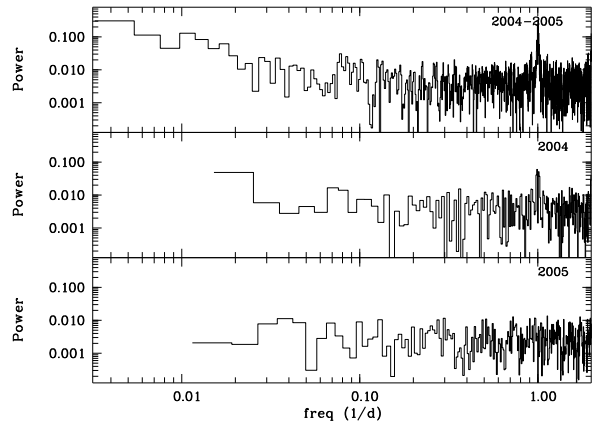}}
\caption{ The power spectrum of the ROTSE-IIId light curve by using Scargle
method. Upper panel is for all time series starting from Sep 2004 to Dec 2005.
Middle and lower panels are  for the 2004 and 2005 data, respectively.}
\label{figlabel}
\end{figure}

\section{Discussion}

Start time of our observations (MJD 53255) corresponds to the orbital phase
 of 0.42 which is calculated using T$_{\pi/2}$ = 51985.31 MJD
from Galloway et al. (2004) (T$_{\pi/2}$ is the epoch at which the mean
longitude is $\pi$/2 ). Maximum of brightness occurs at $\sim$ 0.25
orbital phase. Magnitude  change lasts about 70-80 d.
The brightening and fading phases have similar time durations.
If the magnitude change is attributed
to the disk around Be star then the disk is lost in about 70-80 d.
Hanuschik et al. (1993), in their study of H$\alpha$ outburst of
$\mu$ Cen, propose that the outburst is caused by the ejection of
 photospheric matter into bound orbits forming a short lived (less than
100 d) quasi stable Keplerian disk. They did not find close relation
between H$\alpha$ emission activity and photometric data.
Negueruela et al. (2003) observed almost no change in the strength of the
H$\alpha$ line (equivalent width:$\sim$ 15 A) during their
 one year observation period which begins one month after the 2000-01
X-ray outburst. They argued that if Be star disk was lost then the
strength of H$\alpha$ line should fade away.
It seems that the long period of X-ray activity in 2000-01 period and
likewise remaining relatively constant the strength of H$\alpha$ line
oppose to the possibility of a short lived disk during this period.
It is less likely to have short lived disk in this system.

The circular orbital velocity of X-ray component is 74 km/s according to
the orbital parameters given by Galloway et al (2004).
 Using the orbital parameters of KS 1947+300,  
Galloway  et al., (2004) estimated the mass function 
as f$_{x}$= 1.71 M$_\odot$.   For a neutron star of mass
1.4 M$_\odot$, the mass of optical component is 3.4 M$_\odot$ when
 the inclination angle i=90$^{o}$. 
Since we did not see eclipses either from RXTE/ASM or ROTSE observations, 
we assumed that inclination angles should be less than 
85$^{o}$.
If we assume masses of 15 and 20 M$_\odot$ for the optical component
the inclination angles are found as 28$^{o}$ and 26$^{o}$, respectively.
The corresponding Roche Lobe radii are 73 and 83 R$_\odot$ which
are much larger than the expected radius of B0V star. It is difficult
to explain the X-ray outburst of this system by mass transfer from Be star
through the inner Lagrangian point of Roche Lobe.
In the truncated viscous disk model
 (Okazaki$\&$Negueruela 2001) the tidal torques on neutron star truncates
 the disk effectively in systems with low eccentricity therefore
preventing the development of an extended and steady disk.
It is difficult to explain the behavior of Be disk in this system
with the viscous disk model.

The maximum X-ray luminosity obtained from ASM data (Fig.1) 
during our 2004 observation period
is $\sim$ $5\,10^{36}$ erg/s which is calculated using the definition
in Galloway et al. (2004)
(1 count/s $\sim$$3.6\,10^{-10}$ erg/s/cm$^{2}$ in
the energy range 1.5-12 keV). This luminosity is typical for
TypeI outburst but its duration is longer than the typical
value for TypeI outburst.
 The derived accretion rate on to the neutron star is calculated as
$\sim$ $4\,10^{-10}$ M$_\odot$/yr. 
It is known that
the mass loss rate derived is in the order of $10^{-8}$ M$_\odot$/yr
for Be stars (Waters et al. 1987, Clark et al. 1999).
Clark et al. (1999) find a mass loss rate of 1.5-9$\,10^{-8}$ M$_\odot$/yr
for the Be/X-ray binary A0535+26.
Hanuschik et al. (1993) calculated the typical mass transfer rate during the
H$\alpha$ outburst of $\mu$ Cen (B2IV-Ve star) which lasted a few days
as $\sim$ $4\,10^{-9}$ M$_\odot$/yr. Such equatorial mass loss rates 
from Be star can produce the observed X-ray luminosity for KS 1947+300.

The increase in optical luminosity of the system is calculated $\sim$
$10^{36}$ erg/s using the value of M$_{v}$ =-4.2 (Vacca et al. 1996).
This increase can be explained as a heating of Be star 
equatorial disk by X-rays.
The nearly zero time lag between X-ray and optical light curves suggests
this result. The X-ray emission from the neutron star is absorbed in
the stellar wind and equatorial disk of the Be star and reprocessed into
the optical wavelengths (Kriss et al. 1983, Mendelson and Mazeh, 1991). 

\acknowledgements

This study was supported by TUG (Turkish National Observatory), TUB\.ITAK.
We acknowledge support from The Scientific and Technological Research Council
of Turkey through project 106T040.
ROTSE is a collaboration of Lawrence Livermore National Lab,
Los Alamos National Lab, and the University of Michigan (http://www.rotse.net).


\begin{thebibliography}{}
\bibitem{} Akerlof, C.W., Kehoe, R.L., McKay, T.A., Rykoff, E.S., Smith, D.A., et al.:
               2003, PASP~, 115, 132
\bibitem{} Baykal, A., K{\i}z{\i}lo\u{g}lu, U., K{\i}z{\i}lo\u{g}lu, N.: 2005,
                IBVS, No:5615
\bibitem{} Borozdin, K., Gilfanov, M., Sunyaev, R., Churazov, E., et al.:
    1990, PAZh., 16, 804
\bibitem{} Chakrabarty, D., Towsian, K., Lars, B., Thomas, A.P.: 1995, ApJ~, 446, 826
\bibitem{} Clark, J.S., Lyuty, V.M., Zaitseva, G.V., Larionov, V.M.,
 Larionova, L.M., et al.: 1999, MNRAS~, 302, 167
\bibitem{} Galloway, D.K., Morgan, E.H., Levine, A.M.: 2004, ApJ~, 613, 1164
\bibitem{} Grankin, K.N., Shevchenko, V.S., Yakubov, S.D.: 1991, PAZh., 17, 991
\bibitem{} Goranskij, V.P., Esipov, V.F., Lyutyi, V.M., Shugarov, S.Y.:
      1991, SvAL~,  17, 399 
\bibitem{} Hanuschik, F.W., Dach, S.J., Baudzus, M., Thimm, G.: 1993, A\&A~, 274, 356
\bibitem{} Mendelson, H., Mazeh, T.: 1991, MNRAS~, 250, 373
\bibitem{} K{\i}z{\i}lo\u{g}lu, U.,  K{\i}z{\i}lo\u{g}lu, N., Baykal, A.: 2005,
AJ~, 130, 2766
\bibitem{} Kriss, G.A., Cominsky, L.R., Remillard, R.A., William, G.,
   Thorstensen, J.R.: 1983, ApJ~, 266, 806
\bibitem{} Nequeruela, I., Israel, G.L., Morco, A., Norton, A.J., Speziali, R.:
 2003, A\&A~, 397, 739
\bibitem{} Okazaki, A.T., Nequeruela, I.: 2001, A\&A~, 377, 161
\bibitem{} Oppenheim, A. V., and Schaffer, R. W.: 1975, Digital Signal 
           Processing (Prentice Hall: New Jersey).
\bibitem{} Reig, P., Nequeruela, I., Fabregat, J., et al.: 2004, A\&A~, 421, 673
\bibitem{} Scargle, J.D., 1982, ApJ~, 263, 835
\bibitem{} Tsygankov, S.S., Lutovinov, A.A.: 2005, AstL~, 31, 88
\bibitem{} Vacca, W.D., Garmany, C.D., Shull, J.M.: 1996, ApJ~, 460, 914
\bibitem{} Waters, L.B.F.M., Cote, J., Lamers, H.J.G.L.M.: 1987, A\&A~, 
           185, 206 

\end{thebibliography}
\end{document}